\documentstyle[aps,balanced,graphics,epsf,longtable,array,times,subfigure,epsfig,float,flafter,prl]{revtex}

\begin{document}
\bibliographystyle{prsty}

\title{Synchronization of Spatiotemporal Chaos:\\ 
The regime of coupled Spatiotemporal Intermittency}
\author{
A. Amengual$^1$, 
E. Hern\'andez-Garc\'\i a$^{1,2}$, 
R. Montagne$^{1,2}$\cite{Raul}, 
and 
M. San Miguel$^{1,2}$
}
\address{$^1$Departament de F\'\i sica, Universitat
de les Illes Balears, E-07071 Palma de Mallorca, Spain \\
$^2$ Instituto Mediterr\'aneo de Estudios Avanzados, IMEDEA (CSIC-UIB) ,
 E-07071 Palma de Mallorca, Spain \cite{www}}

\date{\today}
\maketitle

\begin{abstract}

Synchronization of spatiotemporally chaotic extended systems is considered in 
the context of coupled one-dimensional
Complex Ginzburg-Landau equations (CGLE). A regime of 
coupled spatiotemporal intermittency (STI) is identified and described in terms 
of the space-time synchronized chaotic motion of localized structures.
A quantitative measure of synchronization as a function of coupling 
parameter is given through distribution functions and information measures.
The coupled STI regime is shown to dissapear into regular dynamics for 
situations of strong coupling, hence a description in terms of a single
CGLE is not appropiate.
\end{abstract}

\vskip 0.5cm

\begin{twocolumns}

Two issues of high current interest in the general field of nonlinear dynamics
are the quantitative characterization of different regimes of spatiotemporal
complex dynamics
in extended systems \cite{CrossHohenberg2}
 and the synchronization of chaotic oscillators \cite{pecora90}.
 The characterization
of low dimensional chaos is now a mature subject with well established
techniques, including techniques of chaos control.
In this context, the demonstration that the familiar phenomenon of
synchronization
of two regular oscillators \cite{huygens} by a 
weak coupling can 
also be displayed by chaotic
oscillators is an important new idea. This conceptual development has opened a
new avenue of research with interesting practical implications.
Chaos in extended systems is a much less mature subject and many
investigations are still at the level
of classifying different types of behavior. Concepts and methods of
Statistical Mechanics are
commonly invoked in terms of ``phase diagrams'' and
transitions among different ``phases'' of
behavior\cite{chate1,chate2,ciliberto2,montagne96b}. Still, the possibility of 
a synchronized behavior of spatially extended systems
in a spatiotemporal disordered phase is
an appealing idea that we address in this Letter. More specifically we will
consider an extended
one-dimensional system in a chaotic regime known as Spatiotemporal
Intermittency (STI)\cite{chate2} and we will characterize a coupled STI regime.

By synchronization of two chaotic oscillators $O_1$ and $O_2$ it is 
meant in a strict sense that plotting the time series $O_1(t_i)$ 
{\sl vs} $O_2(t_i)$ one obtains a straight line of unit slope. 
For many practical applications, synchronization of chaotic
oscillations calls for an expanded framework and the concept
of ``generalized synchronization'' has been introduced
\cite{rulkov95,kocarev96a} as the appearance of a functional dependence 
between the two time series. In this context we understand here by 
synchronization the situation in which $O_1(t_i)$ becomes a given known 
function of $O_2(t_i)$, while for independent chaotic oscillators 
$O_1(t_i)$ and $O_2(t_i)$ are independent variables. Transferring 
these concepts to spatially extended systems, we search for correlations 
between the space($x_i$)-time($t_j$) series of two variables 
$O_1(x_i,t_j)$ and $O_2(x_i,t_j)$. The synchronization of $O_1$ and $O_2$ 
occurs when these two space-time series become functionally dependent. 
This idea is different from the one much studied in the context of
coupled map models in which the
coupling and emerging correlations are among the local oscillators of which
the spatially extended system is composed. Here we search for 
correlations of two variables at the same space-time
point. 

Our study has been carried out in the context of Complex Ginzburg Landau
Equations (CGLE) which give a prototype example of chaotic behavior in 
extended systems\cite{CrossHohenberg,hohenbergsaarloos}. Our results 
show that the coupling between two complex amplitudes $A_1 \mbox{ and
} A_2$ ($O_1=|A_1|$ and $O_2=|A_2 |$), in a STI regime described below,
establishes spatiotemporal correlations which preserve spatiotemporal chaos 
but lead to a synchronized behavior: Starting from the independent STI
dynamics of $A_1$ and $A_2$, coupling between them leads to a STI regime
dominated by the synchronized chaotic motion of localized structures in space
and time for $A_1$ and $A_2$. An additional effect observed in our model is 
that the coupled STI regime is destroyed for coupling larger than a given 
threshold, so that the two variables remain strongly correlated but each of 
them shows regular dynamics. At this threshold maximal mutual information
and anticorrelation of $|A_1|$ and $|A_2|$ are approached.

The CGLE is the amplitude equation
for a Hopf bifurcation for which the system starts to oscillate with frequency
$\omega_c$ in a spatially homogenous mode. When, in addition, the Hopf 
bifurcation breaks the spatial translation symmetry it identifies
a preferred wavenumber $K_c$. In one-dimensional systems the amplitudes $A_1$
and $A_2$ of the two counterpropagating
traveling waves with frequency $\omega_c$ and wavenumbers $\pm K_c$ becoming
unstable satisfy coupled CGLE of the form
\begin{eqnarray}
\partial_t A_{1,2} & = & \mu A_{1,2} + (1 + i \alpha) \partial_x^2 A_{1,2}
\nonumber \\
& & - (1+i\beta) \left( |A_{1,2}|^2 + \gamma |A_{2,1}|^2
\right)
 A_{1,2} \label{theEqu}
\>.
\end{eqnarray}
Eq. (\ref{theEqu}) is written here in the limit of negligible 
group velocity. In particular, this limit is of interest to describe the 
coupled motion of the two complex components of a Vector CGLE.
In this context, (\ref{theEqu}) is used to describe vectorial transverse
pattern formation in nonlinear optical systems, and $A_{1,2}$
stand for the two independent circularly polarized components
of a vectorial electric field amplitude \cite{maxi95,toni96}. 
The parameter $\mu$ measures the distance to threshold and $\gamma$ is the 
coupling parameter, taken to be a real number.

Homogeneous solutions of Eq. (\ref{theEqu}) are of the form 
\begin{equation}
A_{1,2}(x,t) = Q_{1,2} e^{ i \omega_{1,2} t } \>.
\label{theSol}
\end{equation}
where:
$\omega_{1,2} = - \beta(Q_{1,2}^2 + \gamma Q_{2,1}^2)$. 
For $\gamma=0$, $Q_{1,2}^2 = \mu $, and the two amplitudes satisfy independent
CGLE whose phase diagram has been
studied in much detail in terms of the parameters $\alpha$ and $\beta$
\cite{chate3,montagne96b}. For $\gamma=0$,
solutions of the type (\ref{theSol}), and other plane waves of different
periodicities, are known to be linearly stable below the
Benjamin-Feir (BF) line
($1+\alpha\beta>0$). Above this line regimes of phase and defect chaos occur.
However, for a range of parameters below the BF line 
there is an additional attractor, coexisting with the one of plane waves,
in which the system displays a form of spatiotemporal chaos known as STI.
In this attractor the solution is intermittent in space and time. 
Space--time plots of $|A_1|$ or $|A_2|$ in the STI regime for $\gamma=0$
are qualitatively similar to the ones shown in Fig. \ref{SincrG} (top). 
The question we address here is how the STI regimes of $A_1$ and $A_2$ change 
when the coupling
$\gamma$ is introduced. We first recall that for a weak coupling situation
($\gamma < 1$)
the solution (\ref{theSol}) with $Q_{1,2}^2 = \mu / (1 + \gamma)$ is linearly
stable 
below the same BF-line $1+\alpha\beta>0$ \cite{maxi95} whereas the
solutions with $Q_1=0$ or $Q_2=0$ are unstable. For large coupling
$\gamma>1$ the competition
between the two amplitudes is such that only one of them survives, so that
linearly stable solutions are either
$Q_1=\sqrt{\mu}$, $Q_2=0$, or $Q_2=\sqrt{\mu}$, $Q_1=0$. For the marginal
coupling $\gamma=1$, $Q_1^2 + Q_2^2 = \mu$
and the phase $\chi = \arctan(Q_1/Q_2)$ is arbitrary.
In addition to these ordered states we also find a STI attractor for coupled
CGLE and values
of $\alpha$ and
$\beta$ which are in the STI region of a single CGLE. Changes of such STI
behavior with varying 
$\gamma$ are shown in Fig. \ref{SincrG} \cite{numerics}.

For small coupling ($\gamma<<1$) we observe that $|A_1|$ and $|
A_2|$ follow nearly independent dynamics with the flat grey regions 
in the space-time plot being laminar regions separated
by localized structures that appear, travel and annihilate. 
In the laminar regions configurations close to (\ref{theSol}) with
$Q_1=Q_2$ occur. Disorder occurs via the contamination by localized
structures. These structures have a rather irregular behavior and, 
in a first approach they can be classified in hole--like or
pulse--like\cite{hohenbergsaarloos}. In Fig. \ref{SincrG} 
these hole--like and pulse--like
structures are associated to black and white localized structures 
respectively. It is argued that the domain of parameters in which
STI exists in the limit $\gamma \rightarrow 0$ is determined by the
condition of stability for those localized structures \cite{kramer93}.
As $\gamma$ increases we observe two facts: First, both
$|A_1|$ and $|A_2|$ continue to display STI dynamics although
in larger and slower space-time
scales. Second, and more interesting, is that the dynamics of $|A_1|$
and $|A_2|$ become increasingly correlated. 
This is easily recognized by focusing in the localized structures: 
A black traveling structure in the space-time plot of $|A_1|$
has its corresponding white traveling structure in the space-time plot of
$|A_2|$ and vice versa. This results in laminar states occurring 
in the same region of space-time for $|A_1|$ and $|A_2|$. 
The coupled STI dynamical regime is dominated by localized structures
in which maxima of $|A_1|$ occur always together with minima
of $|A_2|$ and vice versa. In the vicinity of the localized structures,
and emerging from them, there appear travelling wave solutions of 
(\ref{theEqu}) but with different wavenumber for $|A_1|$ and $|A_2|$
so that $|A_1|\neq |A_2|$. Eventually (going beyond the marginal
coupling $\gamma=1$) the STI dynamics is destroyed and 
$|A_1|$ and $|A_2|$ display only laminar
regions, in which either $|A_1|$ or $|A_2|$ vanish, separated by
domain walls. 
\begin{figure}
\resizebox{80mm}{110mm}{\includegraphics{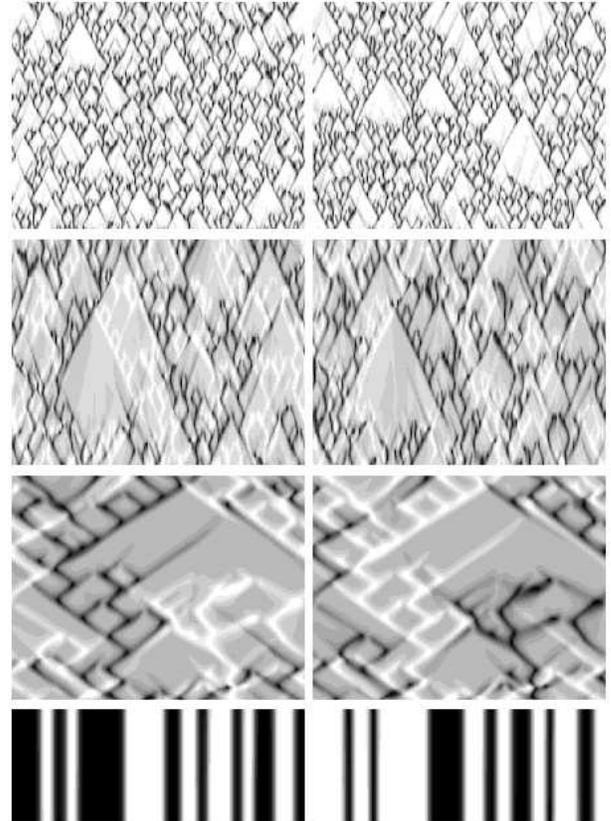}}
\vspace{0.5mm}
\caption{Space-time plot ot the modulus $|A_1|$ (left) and $|
A_2|$ (right) for four values of the coupling parameter $\gamma$:
{}From top to bottom, $\gamma=0.1, 0.5, 0.95$ and $1.05$. The horizontal
axis represents space and the vertical axis time (2000 time
units for $\gamma=0.95$, 100 for $\gamma=1.05$ and 200 in the other two plots).
The grey levels change linearly from the minimum (black) to the maximum (white)
of the modulus.
The parameters are $\mu=1$, $\alpha=0.2$ and $\beta=-2.0$.
For $\gamma=0.1$, the structures appearing in $|A_1|$ and $|A_2|
$ are almost independent.
However, for $\gamma=0.5$, the defects in $|A_1|$
(black and white structures) becomes slightly synchronized with the
defects in $|A_2|$ and, for $\gamma=0.9$, the synchronization is
almost complete.
}
\label{SincrG}
\end{figure}
In the optical interpretation of (\ref{theEqu}) the laminar regions
with $|A_1|= |A_2|$ corresponds to transverse domains
of linearly polarized light, although with a random direction of linear
polarization. The localized structures are essentially circularly
polarized objects since one of the two amplitudes dominates over the other.
Around these structures the plane wave solutions with $|A_1|\neq |A_2|$
have different frequencies, so that they correspond to depolarized
solutions of (\ref{theEqu}) \cite{maxi95}.
As $\gamma>1$, localized traveling structures dissapear and one is left with
circularly polarized domains separated by polarization walls.
\vspace{-4.0mm}
~\begin{figure}[H]
\resizebox{80mm}{70mm}{\includegraphics{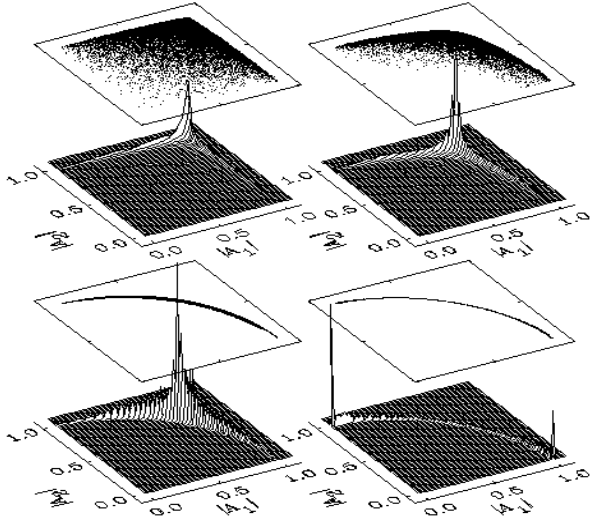}}
\vspace{0.2mm}
\caption{Comparison of the states reached at long times
starting from noise for (from left to right, and from top to bottom)
$\gamma=0.1, 0.5, 0.95$ and $1.05$. The joint probability distribution
$p(|A_2|,|A_1|)$ is shown as a 3D surface. The vertical scale
is arbitrary, the same in the first three plots and
three times larger for $\gamma=1.05$. On top of each surface, 
$|A_1(x,t)|$ vs $|A_2(x,t)|$ are shown in the form of a
dotted plot obtained from the values of $|A_1|$ and $|A_2|$ at space-time
points during a time interval of 50 units.}
\label{LargeT}
\end{figure}
It is usually argued that for $\gamma>1$ the dynamics of the coupled CGLE
(\ref{theEqu}) is well represented by a single CGLE since only one of the two
waves survives. This is certainly not true in 
the STI domain
of parameters considered here since a single CGLE would give rise to STI 
dynamics whereas the coupled
set (\ref{theEqu}) does not for $\gamma>1$. In general a description in terms 
of a single CGLE would not be reliable for parameter values at which the 
single amplitude dynamics produces amplitude values close to zero. 

We next show that the correlations observed for increasing
$\gamma$ in Fig. \ref{SincrG}
are in fact a kind of spatiotemporal synchronization, in the
generalized sense defined in \cite{rulkov95,kocarev96a}. To this end a 
characterization of
the synchronizing process can be given by analyzing the joint distribution of
the two variables. 
This distribution and values of $|A_1|$ versus $|A_2 |$ 
are plotted in Fig.\ref{LargeT}. 
The cloud of points correspond to the different space--time points
of Fig. \ref{SincrG}. 
For $\gamma<<1$ we obtain a
diffuse cloud of points indicating essentially independent dynamics.
The concentration of points
around $|A_1|^2 = |A_2|^2 = \mu/(1+\gamma)$ corresponds to the
laminar regions, but excursions away from that solution
are independent. As the coupling is increased with $\gamma < 1$
the cloud of points approaches the curve given by
$ |A_1|^2 + |A_2|^2 = \mu$. This indicates 
synchronization of the dynamics of structures departing
from the laminar regions. The points with larger values of $|A_1|$ and
smaller values of $|A_2|$ (and vice versa) correspond to the localized 
traveling structures. Intermediate points among
these ones and those around $|A_1|=|A_2|$ correspond to 
the regular solutions of non-zero wavenumber that surround the localized 
structures. The special case of marginal coupling
is discussed below, but as we enter into the strong coupling situation
($\gamma > 1$) the cloud of points concentrates in the regions 
$|A_1|^2=\mu$, $|A_2|=0$ and $|A_2|^2=\mu$, $|A_1|=0$, 
while intermediate points correspond to the domain walls separating these 
ordered regions. It should be pointed out that we are considering just the
modulus of the complex fields $A_{1,2}$. The coupled phase dynamics does not
show synchronization, at least not in an obvious manner, so that we are in 
a case of partial synchronization as considered in \cite{rosenblum96}. 

A quantitative measure of the synchronizing process can be given in terms
of information measures \cite{mceliece}.
The entropy $H(X) = - \sum_x p(x) \ln p(x)$, where $p(x)$ is the
probability that $X$ takes the value $x$, measures the {\sl randomness} of a
discrete random variable $X$.
For two random discrete variables, $X$ and $Y$, with a joint probability
distribution $p(x,y)$, the mutual information
$I(X,Y) = - \sum_{x,y} p(x,y) \ln [p(x) p(y) / p(x,y)] $
gives a measure of the statistical dependence between both variables,
the mutual information being $0$ if and only if $X$ and $Y$ are independent.
Considering the discretized values of $|A_1|$ and $|A_2|$ at space-time 
points as random variables $X=|A_1|$, $Y=|A_2|$, their mutual
information is a measure of their synchronization.
In Fig. \ref{InfMut}(left) we have plotted the mutual information
and the entropy of $|A_1|$ and
$|A_2|$ as a function of $\gamma$ \cite{fn01}.
This graph shows that the entropy of $|A_1|$ and $|A_2|$ 
remains constant for increasing values of $\gamma$, so that
increasing $\gamma$ does not reduce the uncertainty associated with
the single-point distributions of $A_{1,2}$. 
This indicates that synchronization is not here the result of
reduced randomness due to the increase of time and length scales
observed in Fig. \ref{SincrG}. However, the larger is $\gamma$ the larger 
becomes the mutual information, approaching its maximum possible value
($I=H(|A_1|)=H(|A_2|)$) as $\gamma \rightarrow 1$. 
An additional quantitative measurement of synchronization is given
by the linear correlation coefficient 
$\rho = (\langle|A_1||A_2|\rangle - 
\langle|A_1|\rangle\langle|A_2|\rangle)
(var(|A_1|) var(|A_2|))^{-1/2}$ with
$var(x)$ being the variance of $x$.
This coefficient, plotted as a function of $\gamma$ in 
Fig. \ref{InfMut} (right), 
is negative indicating that when $|A_1|$ increases,
$|A_2|$ decreases, and vice versa.

Our quantitative indicators of synchronization, $I$ and $\rho$, approach
their maximum absolute values as $\gamma \rightarrow 1$. We also observe
that the regime of coupled STI disappears for $\gamma > 1$. In fact,
the generalized synchronization observed, manifested by the tendency of the 
space-time signals towards the functional relation 
$|A_1|^2 + |A_2|^2 = \mu$, is probably related to 
the fact that for $\gamma=1$ this is an attracting manifold for homogeneous
states: Writing (\ref{theEqu}) in terms of 
$R^2 \equiv |A_1|^2 + |A_2|^2$ and 
$\chi \equiv \arctan(|A_1|/|A_2|)$,
it is immediate to see that homogeneous solutions for $\gamma=1$ are
$R^2 = \mu$ and $\chi$ arbitrary. To understand the preference for these
solutions it is instructive to look at the transient dynamics starting
from random initial conditions: $R(x,t)$ has a very fast 
evolution towards $R= \sqrt \mu$ with no regime
of STI existing at any time. During this fast evolution, the phase $\chi(x,t)$ 
covers almost completely the range of its possible values.
The late stages of the dynamics are characterized by a spatial diffusion
of the phase $\chi(x,t)$ until it reaches a space-independent arbitrary
value $\chi(x,t)=\chi_0$.
In a $|A_1|$ {\sl vs} $|A_2|$ dotted plot as the ones in Fig. \ref{LargeT}
this is visualized by a cloud of points quickly approaching $R^2=\mu$ and
then collapsing into a single point. Runs with different random initial conditions
lead to different $\chi_o$. An underlying reason for the special dynamical
behavior at $\gamma=1$ is the separation of time scales for $R$ and $\chi$.
For $\gamma \neq 1$, the zero wavenumber components of $R$ and $\chi$ have
a nonzero driving force and they compete dynamically but, at $\gamma=1$,
$\chi(k=0)$ is a marginal variable, while $R(k=0)$ is strongly driven.
As a consequence, $R$ relaxes quickly towards $R^2=\mu$. Once $R$
becomes space-homogeneous, the different wavenumber components of 
$A_{1,2}$ are decoupled and the zerowavenumber solution wins by
diffusion of $\chi$.
\begin{figure}[H]
\vspace{-8mm}
\mbox{
\hspace{-12mm}
\subfigure{\resizebox{50mm}{42mm}{\includegraphics{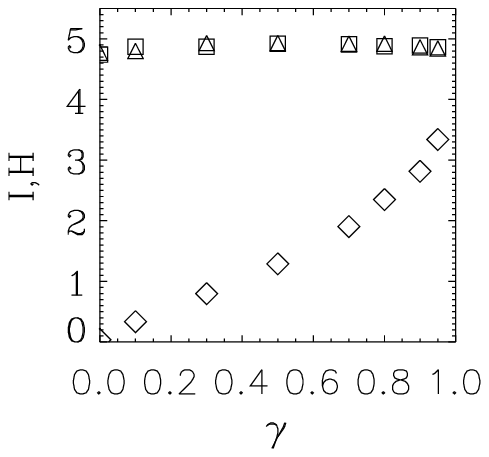}}}
\hspace{-10mm}
\subfigure{\resizebox{50mm}{42mm}{\includegraphics{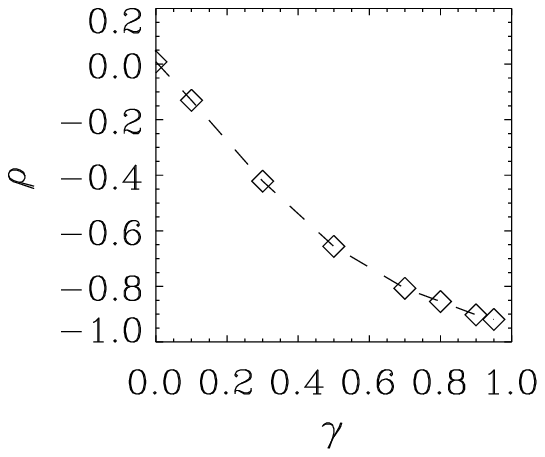}}}
}
\vspace{-5mm}
\caption{
Left: Entropy of $|A_1|$ ($\Box$) and $|A_2|$ ($\bigtriangleup$) 
and their mutual information I ($\diamond$) as a function of $\gamma$.
Right: Correlation coefficient $\rho$ of $|A_1|$ {\sl vs} $|A_2|$
as a function of $\gamma$.}
\label{InfMut}
\end{figure}
In some of our simulations the STI regime has been observed to disappear for a 
coupling smaller than $\gamma=1$, but this seems to be a consequence of 
finite-size effects: 
The size of the laminar portions of Fig.(\ref{SincrG}) increases with 
the coupling $\gamma$. 
When this size becomes similar to system size, one of the stable
plane waves can occupy the whole system, thus preventing any further
appearance of defects and STI. For a given initial condition, with
parameters $\alpha=0.2$ and $\beta=-1.4$, and a system size $L=512$
the STI regime was seen to disappear at $\gamma=0.85$. As soon as 
the system size was doubled the STI regime reappeared again. By reducing 
system size to $L=256$ the STI regime disappeared for smaller $\gamma$. The
conclusion from this an other numerical experiments is that STI exists for 
all $\gamma<1$ in the same range of parameters as it exists in the single CGLE,
with time and length scales diverging as $\gamma$ approaches $1$, where STI
disappears. 

In summary we have described a regime of synchronized STI dominated by
the space-time synchronization of localized structures. Synchronization
is measured by a mutual information and a correlation parameter that take
their absolute maximum value at the boundary between weak and strong coupling
$\gamma=1$. Beyond this boundary ($\gamma>1$) STI disappears but the strong,
coupled system dynamics cannot be described in terms of a single dominant
amplitude.

Financial support from DGICYT Project PB94-1167 (Spain) and European Union
Project QSTRUCT (FMRX-CT96-0077) are acknowledged. R.M. also acknowledges 
partial support from the PEDECIBA and CONICYT (Uruguay).

\vskip -0.3cm



\end{twocolumns}

\begin{thebibliography}{10}
\bibitem{numerics}
Our numerical integration is performed by a second order pseudospectral method
  described elsewhere \cite{montagne96c}. The numerical results presented
  in this paper have been obtained, except when indicated explicitly, 
  for $L=512$, and, without
  loss of generality, $\mu=1$. The initial condition is white Gaussian noise with a mean
  squared amplitude at each point of $0.002$. In Fig. \ref{SincrG}, the same 
  initial condition is taken for all values of $\gamma$.
\bibitem{fn01}
To calculate $H$ and $I$ we have discretized the range of possible values $0 <
  |A_i| < 1.1$ in $N=200$ bins. Therefore, the maximum possible entropy is $\ln
  200 = 5.3$.
\end{thebibliography}
\end{document}